\documentclass[aps,pra,reprint, amsmath, amssymb,superscriptaddress]{revtex4-1}
 
\usepackage{bm}
\usepackage[retainorgcmds]{IEEEtrantools}
\usepackage{graphicx}
\usepackage{mathrsfs}
\usepackage{amsmath}
\usepackage{amsfonts}
\usepackage{amssymb}
\usepackage{color}
\usepackage{times,txfonts}
\usepackage{nicefrac}
\usepackage{ragged2e}
\usepackage{tikz}

\usepackage[colorlinks=true,linkcolor=blue,urlcolor=blue,citecolor=blue,pdfusetitle]{hyperref}

\usepackage{blindtext}

\definecolor{cadmiumgreen}{HTML}{097969}

\DeclareMathOperator*{\argmax}{arg\,max}
\DeclareMathOperator{\EX}{\mathbb{E}}

\begin{document}

\title{Bayesian estimation for collisional thermometry}
\date{\today}
\author{Gabriel O. Alves}
\email{alves.go.co@gmail.com}
\affiliation{Instituto de F\'isica da Universidade de S\~ao Paulo,  05314-970 S\~ao Paulo, Brazil.}
\author{Gabriel T. Landi}
\email{gtlandi@gmail.com}
\affiliation{Instituto de F\'isica da Universidade de S\~ao Paulo,  05314-970 S\~ao Paulo, Brazil.}

\begin{abstract}
Quantum thermometry exploits the high level of control in coherent devices to offer enhanced precision for temperature estimation.
This highlights the need for constructing concrete estimation strategies.
Of particular importance is collisional thermometry, where a series of ancillas are sent sequentially  to probe the system's temperature. 
In this paper we put forth a complete framework for analyzing collisional thermometry using Bayesian inference. 
The approach is easily implementable and  experimentally friendly.
Moreover, it is guaranteed to always saturate the Cramér-Rao bound in the long-time limit. 
Subtleties concerning the prior information about the system's temperature are also discussed, and analyzed in terms of a modified Cramér-Rao bound associated to van Trees and Sch\"utzenberger.

\end{abstract}

\maketitle{}

\section{Introduction}

% Dar um double check nesses papers

Recently, there has been considerable progress in our understanding of the ultimate bounds on thermometric precision.  Using tools from quantum parameter estimation, Refs.~\cite{Mitchison2020, DePasquale2017, Jevtic2015, Seah2019, Mehboudi2018, Salado-Mejia2021, Correa2015, Razavian2019, Mukherjee2019, Planella2020, Correa2017, Hovhannisyan2018} have put forth several case studies of optimal thermometry in the quantum regime. 
Within this context, the concept of optimality is typically quantified through the Quantum Fisher Information (QFI), which establishes the Cramér-Rao bound (CRB), a lower bound for the variance of unbiased estimators.
By maximizing the QFI one can thus also improve the limits of precision of a given estimation \cite{PARIS2009}.
These analyses have the advantage of being independent of the actual estimators being used to assess the temperature. They therefore provide a global view on the problem.  
However, they lack the concrete prospect of practical implementations in the laboratory. 

\begin{figure}[htp]
  \centering
  \includegraphics[width = \columnwidth]{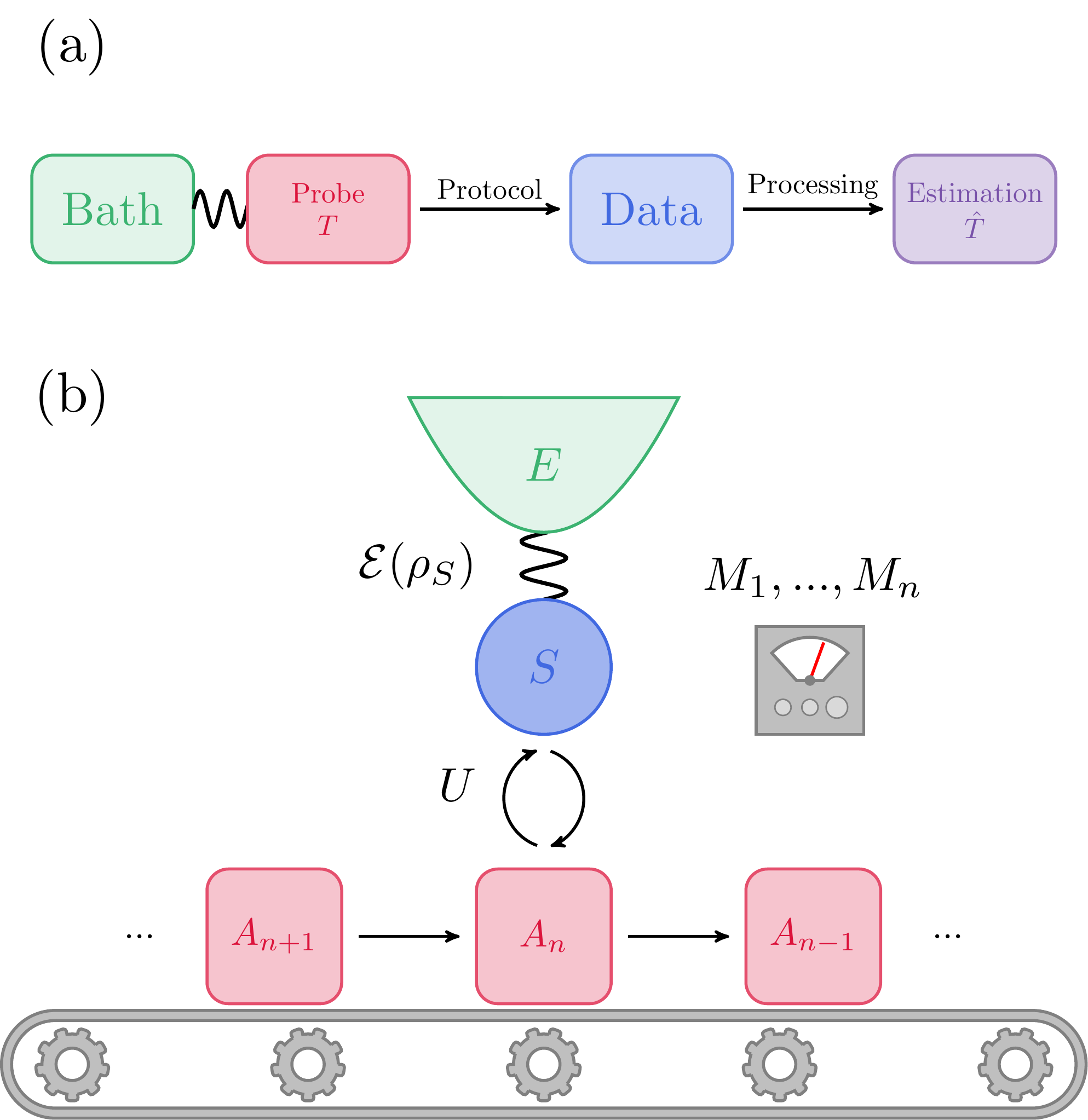}
  \caption{(a) Diagram depicting the typical procedure in probe-based thermometry. Information about the temperature of a bath is extracted via a probe. 
  Most studies focus on how to maximize the information contained in the extracted data.
  A less explored, but equally important problem, is how to actually \emph{process} that data into concrete estimators. (b) An illustration of the collisional model, which will be used in this paper to discuss estimators based on Bayesian Inference. 
  }
  \label{fig:diagram}
\end{figure}

This concerns a different, more practical challenge: how to construct concrete thermometry protocols, specially when it comes to data processing. As depicted in Fig.~\ref{fig:diagram} (a),
this is a complementary, and thus ultimately different, task.

The maximum likelihood estimator (MLE) is one of the canonical choices in this sense. Its desirable asymptotic properties, such as unbiasedness, make it one of the standard choices in parameter estimation \cite{miller_miller_freund_2019, Fiurasek2001, Suzuki2020, Ly2017}. However, certain minimization criteria \cite{Kay1993}, symmetries and constraints in the problem may enforce different choices for estimators and figures of merit, such as in \cite{Escher2011, VonToussaint2011, Rubio2020}. In this sense, a variety of authors have recently studied more concrete implementations of estimators \cite{Morelli2021, Li2018, Kiilerich2016, Teklu2009, Hanamura2021,Rubio2020}. 

An interesting platform for temperature estimation is that of collisional thermometry (Fig.~\ref{fig:diagram} (b)), put forth in~\cite{Seah2019} (see also \cite{Shu2020}).
In standard probe-based thermometry, a series of ancillas $A_n$ are sent to sequentially probe the temperature of a certain system of interest, $E$. 
Usually one assumes that $E$ is sufficiently large, so as to not be degraded by the contact with the ancillas. 
Conversely, in collisional thermometry an intermediate system $S$ is placed in between $A_n$ and $E$. 
Since the ancillas are never allowed to thermalize with $E$, the problem is intrinsically out-of-equilibrium. 
This introduces (at least) two advantages. 
First, it encodes information about the temperature in the dynamical relaxation rates of the probe, allowing one to exceed the maximal precision that would be possible if $S$ was not present (the so-called Thermal Fisher Information). 
Second, it creates correlations between the ancillas which, together with collective measurements, can be used to obtain an additional boost in precision.

Collisional thermometry is a scalable platform, where statistics from an arbitrary number of ancillas can be accumulated to obtain increasingly higher  precision. 
However, as with most other thermometry schemes, there has so far been no studies discussing concrete estimators for it. 
That is, once the data is obtained, how do we actually use it to infer the temperature $T$?
The goal of this paper is to fill in this gap.
We show that Bayesian estimation (BE) provides a  powerful tool-set for thermometry, which is both easy to implement and experimentally friendly. 
BE has already been extensively employed in  quantum metrology \cite{Morelli2021, Li2018, Kiilerich2016, Teklu2009} and open quantum systems \cite{Kiilerich2015, Zhang2017, Gammelmark2013}. BE has also been used in thermometry before~\cite{Rubio2020, Jorgensen2021, Boeyens2021}, first appearing in \cite{Prosper1993}. Through sequential measurements one uses Bayes rule to continuously update and refine the state-of-knowledge about the parameter's distribution. 
By doing so we are able to construct  estimators which can be used to infer the temperature of a reservoir. 

We focus  on the so-called Bayesian Average (BA), which minimizes the Bayesian mean-squared-error (BMSE) \cite{Berger1985}. 
The latter can be used to provide a concise evaluation of the estimator's performance for a wide-range of temperatures, a desirable property for thermometers. 
Moreover, it also avoids two conundrums that are sometimes found in other strategies.
First, it does not require unbiased estimators,
which can sometimes be unphysical \cite{Escher2011}, or impractical (c.f. chapter 2 of~\cite{Kay1993} for a detailed discussion).
% (for instance, when the estimator which minimizes the variance is not unique for the whole parameter space \footnote{The adoption of certain minimization criteria, such as the typical mean-squared error minimization, may lead to unrealizable estimators. One instance is when the estimator depends not only on the data but also on the very parameter being estimated. Constraining the bias to zero is an alternative which also typically solves this problem, however the tradeoff is that there may not exist a single estimator which minimizes the variance for the whole parameter space. One unbiased estimator might be optimal for one parameter but not necessarily for the others. For more details see \cite{Kay1993}.}), 
Although the BA is biased, this always vanishes asymptotically for a large number of ancillas. 
Second, the BMSE provides a measure of precision averaged over the entire range of temperature of interest, unlike the standard CRB, which requires knowledge of the very temperature one is trying to estimate. 
The BMSE can also be compared with a bound by van Trees and Sch\"utzenberger \cite{VanTrees2001, Schutzenberger1957, bj/1186078362}, which represents a Bayesian analogue of the CRB.

The paper is divided as follows.
In Sec.~\ref{sec:collisional_thermometry} we briefly review the main results of collisional thermometry, based on Ref.~\cite{Seah2019}, which will be the model used in this work.
In Sec.~\ref{sec:bayesian_estimation} we discuss  techniques from Bayesian estimation, which are then applied in   Sec.~\ref{sec:analysis}.
Final remarks and future prospects are discussed in Sec.~\ref{sec:discussions_and_conclusions}.

\section{\label{sec:collisional_thermometry}Collisional Thermometry}

\subsection{Standard probe-based thermometry}

In standard probe-based thermometry, ancillas are sent to interact directly with a reservoir $E$, kept at a fixed temperature $T$. 
After this interaction they will be in a certain state $\rho_A(T)$, which contains  information about $T$ that must be extracted via some measurement strategy. 
The error $\epsilon$ in any unbiased temperature estimator, constructed from this measurement, is lower bounded by the Cramér-Rao bound (CRB)
\begin{equation}\label{cramer_rao}
    \epsilon \geqslant \frac{1}{nF(T)},
\end{equation}
where $n$ is the number of measurement outcomes, and $F$ is the Fisher Information (FI) associated to the state $\rho_A(T)$ and the measurement strategy employed.
Given measurement outcomes $p_x(T)$, the latter is defined as 
\begin{equation}\label{FI_standard}
    F(T) = \sum\limits_x  \frac{1}{p_x}\left( \frac{\partial p_x}{\partial T} \right)^2.
\end{equation}
The FI maximized over all possible measurement strategies, is known as the Quantum Fisher Information (QFI), and is given by
\begin{equation}\label{QFI_standard}
    \mathcal{F} = {\rm tr}\big\{\Lambda^2 \rho_A\big\},
\end{equation}
where $\Lambda$ is the symmetric logarithmic derivative (SLD), which is a solution of $\Lambda \rho_A + \rho_A \Lambda = 2 \partial_T \rho_A$. 

The optimal scenario occurs when the probe fully thermalizes with the environment~\cite{Jevtic2015, Correa2015, Liu2020}. 
That is, when $\rho_A$ is a thermal state $\rho_A^{\rm th} = e^{-\beta H_A}/Z_A$, with  $Z_A=\text{tr}(e^{-\beta H_A})$ and $\beta = 1/T$.
In this case, the QFI reduces to the thermal Fisher information 
\begin{equation}\label{TFI}
\mathcal{F}_{th} = \frac{C}{T^2}, \hspace{15pt} 
 C = \frac{\langle H_A^2 \rangle -  \langle H_A \rangle^2}{T^2},
\end{equation}
where $C$ is the ancilla's heat capacity.

\subsection{Collisional thermometry}

In this paper we focus instead on collisional thermometry (Fig.~ \ref{fig:diagram}(b)), which represents a generalization of the scenario above. 
A finite system $S$ is placed between $A_n$ and $E$, thus serving as an indirect connection between them. 
The interactions are piecewise and alternating: first the system interacts with $E$ for a certain time $\tau_{SE}$. Then they are decoupled and the system interacts with the ancilla, for a certain time $\tau_{SA}$. 
The process is then repeated, each time with a new ancilla.

For concreteness, we take both $S$ and all the $A_n$ to be resonant qubits, with $H_S = \Omega \sigma_Z^S/2$ and $H_{A_n} = \Omega \sigma_Z^{A_n}/2$, where $\sigma_Z$ are Pauli matrices. The system interacts with $E$ through the a quantum master equation, which introduces a temperature dependence on the state of the system:
\begin{equation}
\frac{d\rho_S}{dt} = \mathcal{L}(\rho_S) = \gamma(\bar{n} + 1)\mathcal{D}[\sigma_-^S] + \gamma \bar{n} \mathcal{D}[\sigma_+^S],
\end{equation}
where $\mathcal{D}[L] = L\rho L^\dagger - \frac{1}{2}\{L^\dagger L, \rho \}$, $\gamma$ is the coupling strength and $\bar{n} = 1/(e^{\Omega/T} - 1)$ is the Bose-Einstein occupation. 
The $SE$ interaction is thus described by the map $\mathcal{E}(\rho_S) = e^{\tau_{SE} \mathcal{L}}(\rho_S)$.
Conversely, the system-ancilla interaction is chosen to be a partial-swap \cite{Scarani2002}: 

\begin{equation}
    U_{SA_n} = \exp\Big\{ - i \tau_{SA} g(\sigma_+^S \sigma_-^{A_n} + \sigma_-^S \sigma_+^{A_n})\Big\}.
\end{equation}
All ancillas are assumed to start in the same initial state $\rho_A^0$, which we take to be the ground-state $\rho_A^0 = |0\rangle\langle 0|$. 
The coupling strengths $\gamma \tau_{SE}$ and $g \tau_{SA}$ are thus the free parameters of our model.
And temperature is measured throughout in units of $\Omega = 1$.

From the perspective of the system, the alternating application of these two maps yields a stroboscopic evolution,
\begin{equation}\label{eq:stroboscopic_map}
\rho_S^n =  \text{tr}_{A_n}\{\mathcal{U}_{SA_n} \circ \mathcal{E}(\rho_S^{n-1} \otimes \rho_A^0)\} := \Phi(\rho_S^{n-1}),
\end{equation}
where $n = 1,2,3,\ldots$ labels the collisions. 
Here $\mathcal{U}_{SA_n}(\bullet) = U_{SA_n} \bullet U_{SA_n}^\dagger$ and $\circ$ denotes map composition.
We always consider steady-state operation regimes. That is, we first allow several ancillas to collide with the system, so that it reaches a fixed point $\rho_S^* = \Phi(\rho_S^*)$. 
This eliminates any transient effects, making problem  translationally invariant, which is highly advantageous. 

\begin{figure}
     \centering
     \includegraphics[width=\columnwidth]{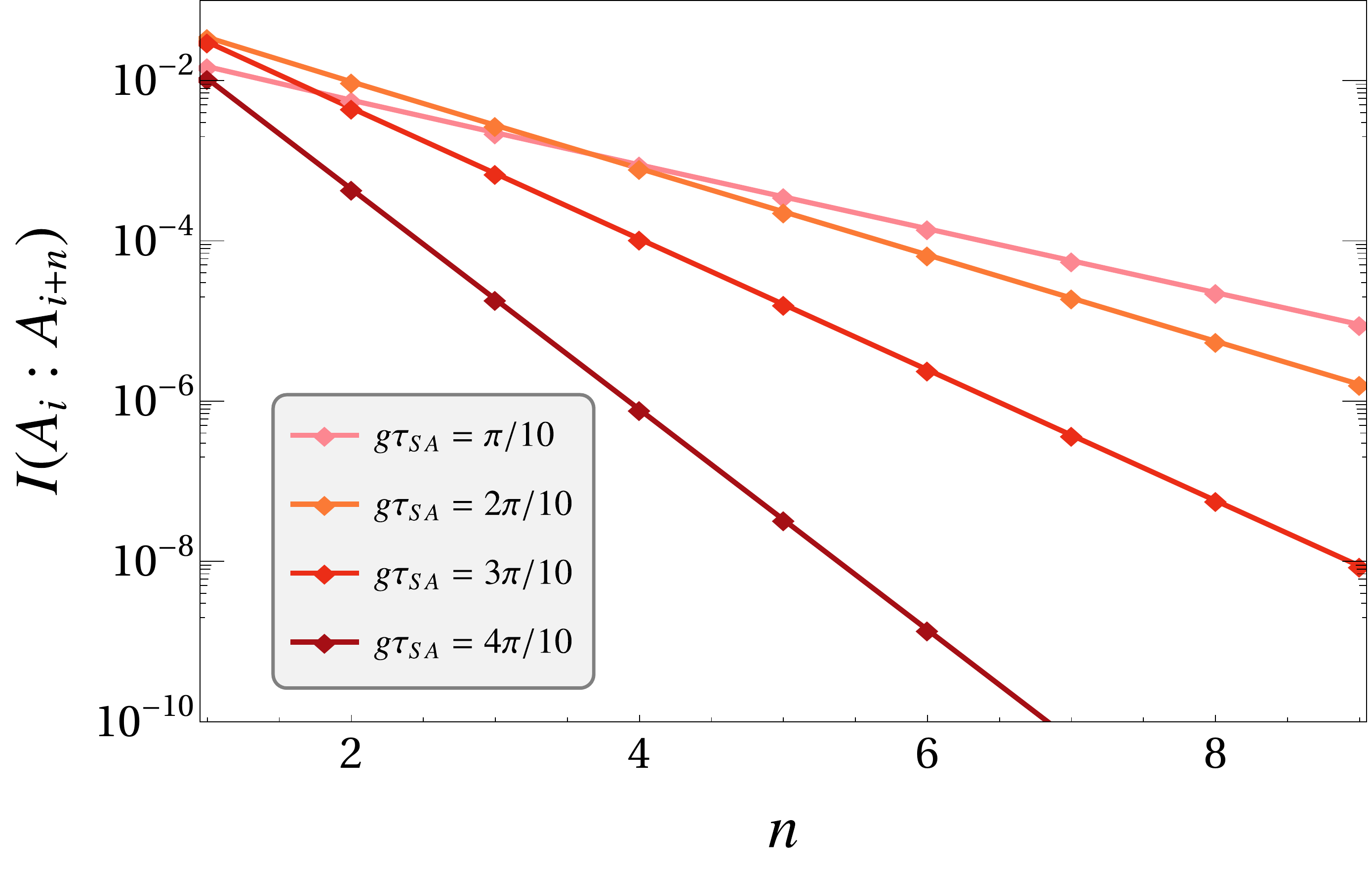}
     \caption{A log-linear plot of the correlation between an ancilla and its $n$-th neighbor for different values of SA coupling. Here we consider $T/\Omega = 2$ and $\gamma \tau_{SE} = 0.2$.
     }
     \label{fig:mutualinfodecay}
\end{figure}

\subsection{Decay of ancilla-ancilla correlations}

From the perspective of the ancillas, the presence of $S$ will cause them to become correlated with each other. 
Starting from $\rho_S^*$, the joint state for a block of ancillas $\rho_{A_i \ldots A_{i+n}}$ will be given by 
\begin{equation}
    \rho_{A_i \ldots A_{i+n}} = {\rm tr}_S \Big\{ \mathcal{U}_{SA_{i+n}} \circ \mathcal{E} \circ \ldots \circ \mathcal{U}_{SA_i} \circ \mathcal{E} \big(\rho_S^* \otimes \rho_A^0 \otimes \ldots \otimes \rho_A^0 \big)\Big\}.
\end{equation}
Since $\rho_S^*$ is a steady-state,  $\rho_{A_i \ldots A_{i+n}}$ will be translationally invariant; that is, independent of $i$. 

Information about $T$ is extracted from $\rho_{A_i \ldots A_{i+n}}$ by performing a measurement on the ancillas, described by a positive operator valued measure (POVM). 
Since this global state is correlated, several choices of measurement strategies arise. 
In fact, as shown in~\cite{Seah2019}, these correlations can actually be used to further enhance the precision. 
However, this requires collective POVMs, which are hard to implement. 
For concreteness, we will focus here only on local measurements. 
We let $\{M_x\}$ denote a set of POVM elements acting on a single ancilla, with possible outcomes $x=0,1$.
The joint distribution obtained from measuring a block of $n$ ancillas will then be 
\begin{equation}\label{likelihood_geral}
    P(X_n,\ldots,X_1|T) = {\rm tr} \Big\{ M_{X_n} \ldots M_{X_1} \rho_{A_1 \ldots A_{n}}\Big\}.
\end{equation}

For the correlations to be significant, some fine tuning of the parameters is required; e.g. taking very small  interaction times $\tau_{SA}$ or very low temperatures. 
To quantify this, we consider the mutual information  between any pair of ancillas $A_i$ and $A_{i+n}$: 
\begin{equation}
I(A_i{:}A_{i+n}) = S(\rho_{A_i}) + S(\rho_{A_{i+n}}) - S(\rho_{A_i A_{i+n}}),
\end{equation}
where $S(\rho) = - {\rm tr}(\rho \ln \rho)$ is the von Neumann entropy. 
Fig.~\ref{fig:mutualinfodecay} shows $I(A_i{:}A_{i+n})$ (which is independent of $i$) as a function of $n$, for typical parameters. 
As can be seen, the correlations decay exponentially with distance, and are quite small, already for nearest-neighbors, $I(A_i{:}A_{i+1})$.

Restricting to only local POVM places further restrictions on how these correlations can be accessed. 
As a consequence, to a good approximation one may take the outcomes to be independent and identically distributed (iid). That is, 
\begin{equation}\label{eq:DistributionApproximation}
P(X_n, ..., X_1|T) \approx P(X_n|T)...P(X_1|T),
\end{equation}
where 
\begin{equation}\label{eq:CollisionalLikelihood}
P(X_i|T) = \text{tr}(M_{X_i} \rho_{A_{i}}).
\end{equation}
It should be stressed, however, that the BE formalism that will be described in Sec.~\ref{sec:bayesian_estimation} does not require this assumption; it simply facilitates the analysis. 
In fact, in appendix~\ref{app:corr} we discuss how to extend all results  to the case when~\eqref{eq:DistributionApproximation} is no longer satisfied.

\subsection{Single ancilla QFI}

The optimal choice of POVM is determined by computing the SLD and the QFI in Eq.~\eqref{QFI_standard}~\cite{PARIS2009}. 
Due to our choices of initial ancilla state, and SA interaction, the states $\rho_{A_i}$ are diagonal and hence the optimal measurement is just a projective measurement in the computational basis,
$M_1 = | 0 \rangle \langle 0 |$ and $M_2 = | 1 \rangle \langle 1 |$.
This yields populations $p_1 = P(X_i = 1| T)$ and $p_0 = 1- p_1$. 

\begin{figure}[b]
     \centering
    \includegraphics[width=\columnwidth]{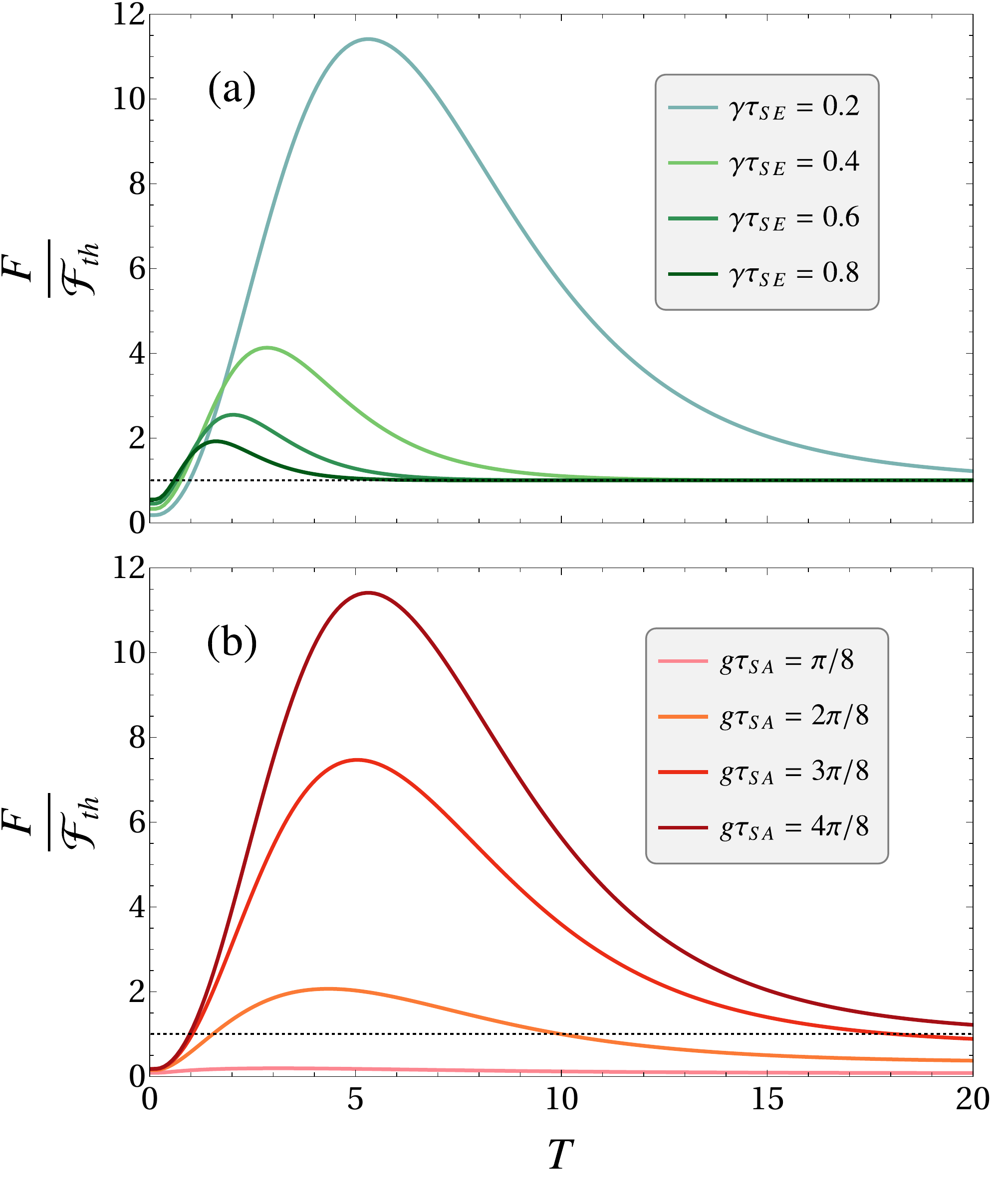}
     \caption{Ratio between the Fisher information and the thermal Fisher information for (a) full swaps, with $g \tau_{SA} = \pi/2$, and for (b) $\gamma \tau_{SE} = 0.2$. The dotted line highlights the ratio $F/\mathcal{F}_{th} = 1$.
     }
     \label{fig:FisherComparison}
\end{figure}

The corresponding Fisher information~\eqref{FI_standard},  of each ancilla, is now readily found to be 
\begin{equation}\label{eq:fisher}
F = \sum_{i= 0, 1} \frac{1}{p_i}\left( \frac{\partial p_i}{\partial T} \right)^2 = \frac{1}{p_1(1-p_1)} \left( \frac{\partial p_1}{\partial T} \right)^2.
\end{equation}
In this case the FI is also the QFI~\eqref{QFI_standard} (i.e. the measurement is optimal). 
But the framework developed in the next section equally holds for a generic FI, not necessarily the QFI, so we shall henceforth continue to write this as $F$, instead of $\mathcal{F}$. 
For comparison, the thermal Fisher information~\eqref{TFI}, which would be obtained if the ancillas fully thermalized with the bath, reads  
\begin{equation}\label{qubit_TFI}
    \mathcal{F}_{\rm th} = (\Omega/2T^2)^2 {\rm sech}^2(\Omega/2T).
\end{equation}
In Fig.~\ref{fig:FisherComparison} we plot the ratio between $F/\mathcal{F}_{\rm th}$ as a function of temperature for different parameters. 
As can be seen, the ratio can be well above unity, showing that the collisional thermometry protocol can offer  significant improvements over  standard probe-based thermometry, for a wide range of temperatures. 
\section{\label{sec:bayesian_estimation}Bayesian Estimation}

Having described the basic model, we now turn to Bayesian estimation (BE) as the basic tool for constructing concrete estimators. 
BE is centered around two main ideas. First, that even though the true temperature $T_0$ is not  known, we still have some prior information about it, which can be used to aid the estimation.  This is done by treating $T$ as a random variable, with  whatever we previously know about it condensed in a distribution $P(T)$, called the prior.
In thermometry, making explicit use of such prior information is crucial:  temperature can in principle vary over enormous scales, but almost always one knows that it lies within a well defined interval. 
For example, one can say with certainty that the temperature of a Bose-Einstein condensate is not 10 K. 
In fact, for many experiments, including Bose gases, said intervals can be very narrow~\cite{Olf2015}. 
The issue of how to quantify such narrowness is discussed in Refs.~\cite{Rubio2020,Mok2021}. 

% (for an example of typical experimental parameters, see \cite{Olf2015,Mehboudi2019}. A quantitative and rigorous approach regarding the local-global dichotomy can be found in \cite{Rubio2020,Mok2021}).}

Second, once the measurement outcomes $\bm{X} = (X_1, ..., X_n)$ are obtained,
one should update the prior with this new information.
This leads to the so-called posterior distribution, which is determined by Bayes' theorem:
\begin{equation}\label{eq:BayesTheorem}
P(T|\boldsymbol{X}) = \frac{P(\boldsymbol{X}|T)P(T)}{P(\boldsymbol{X})},
\end{equation}
where $P(\bm{X})= \int P(\boldsymbol{X}|T)P(T) dT$ and $P(\bm{X}|T)$ is given by Eq.~\eqref{eq:DistributionApproximation}.
Since this refers to independent outcomes, the Bernstein-von Mises theorem~\cite{Butler2007, Cam1986, Vaart1998} ensures that the posterior will converge, in the limit of large $n$, to a Gaussian with mean $T_0$ (the real parameter) and variance $1/nF(T_0)$, where $F$ is the Fisher information. 
In symbols
\begin{equation}\label{posterior_gaussian}
P(T|\boldsymbol{X}) \approx \sqrt{\frac{n F(T_0)}{2\pi}}e^{-\frac{nF_0(T-T_0)^2}{2}},
\quad (n~\text{large}).
\end{equation}
Hence, within this collisional thermometry setting, Bayesian estimation is guaranteed to converge to the true value, with a variance that saturates the CRB~\eqref{cramer_rao}.
This is highly advantageous.
One should also emphasize that the framework is not restricted to independent outcomes, and may also efficiently be implemented for generic $P(\bm{X}|T)$, as discussed in Appendix~\ref{app:corr}.

\subsection{\label{subsec:BayesianEstimators}Estimators}

One of the most widely used estimators in this context is the Bayesian Average (BA):
\begin{equation}\label{eq:estimator}
\hat{T}(\boldsymbol{X}) = \int T P(T|\boldsymbol{X}) dT 
\end{equation}
This can be shown to minimize the Bayesian mean-squared  error (BMSE)~\cite{VanTrees2001}
\begin{equation}\label{eq:BayesianRisk1} 
\epsilon_{B}(\hat{T}(\boldsymbol{X}))
=
\int P(T)dT
\int(T - \hat{T})^2P(\boldsymbol{X}|T)d\boldsymbol{X}
\end{equation}
Moreover, due to Eq.~\eqref{posterior_gaussian}, it is guaranteed to converge to the true value of the parameter in the large $n$ limit. 
Another common choice of estimator \footnote{More generally, one may start with a generic cost function $C(\hat{T},T)$ instead of  $(\hat{T}-T)^2$ in Eq.~\eqref{eq:BayesianRisk1}. Different estimators can then be constructed as those which minimize the corresponding cost function.}
is the maximum a posteriori (MAP), given by $\hat{T}_{MAP} (\boldsymbol{X}) = \argmax_{T} P(T|\boldsymbol{X})$. i.e. by the mode of the posterior. 
The BA, however, has nicer general properties, and is also very easy to compute. We will henceforth focus solely on it, for concreteness. 

A crucial difference, with respect to standard parameter estimation lies in the fact that the error~\eqref{eq:BayesianRisk1} is  averaged over the prior $P(T)$, since $T$ is treated as a random variable.
This can be compared with the usual mean-squared error, which is defined as 
\begin{equation}\label{eq:FrequentistRisk1} 
\epsilon(\hat{T}(\boldsymbol{X})|T)
=
\int(T - \hat{T})^2P(\boldsymbol{X}|T)d\boldsymbol{X}.
\end{equation}
This is, for instance, the quantity appearing in the CRB~\eqref{cramer_rao}. 
As can be seen, it  is conditioned on the value of $T$ (averaged solely over different realizations of the data); we shall henceforth refer to it simply as the mean-squared error (MSE).
The two quantities are connected by 
\begin{equation}\label{eq:FrequentistRisk2} 
\epsilon_{B}(\hat{T}(\boldsymbol{X}))
=
\int 
\epsilon(\hat{T}(\boldsymbol{X})|T)
P(T)
dT.
\end{equation}
The Bayesian error~\eqref{eq:BayesianRisk1} therefore represents a figure of merit which does not depend on the particular value of the parameter. 
This is interesting since the parameter is not known in the first place. Hence, it provides a  way of assessing the \emph{overall} performance of a thermometric protocol, averaged over the prior information.

A third figure of merit, which is of interest in experimental settings, is the \emph{posterior loss}. In this case, one is interested in the error with respect to a particular realization $\bm{X}$. We then calculate it as $\epsilon_{\bm{X}} = \int (T - \hat{T})^2P(T|\boldsymbol{X})dT$
. When the estimator $\hat{T}(\bm{X})$ is chosen to be the posterior mean, the posterior loss $\epsilon_{\bm{X}}$ can be interpreted simply as the variance of the posterior distribution $P(T|\bm{X})$.

Unbiased estimators satisfy the CRB in Eq.~\eqref{cramer_rao}. They, however, have two disadvantages: (i) they depend on the value of the parameter we are trying to estimate, which is not known; and (ii)  they do not take into account any prior information.

Both of these issues are  taken into account by the BMSE~\eqref{eq:BayesianRisk1}. 
Instead of the CRB, this error satisfies the van Trees-Sch\"utzenberger bound (VTSB)~\cite{VanTrees2001, Schutzenberger1957, bj/1186078362}
\begin{equation}\label{eq:VanTrees}
\epsilon_{B}(\hat{T}(\bm{X})) \geq \frac{1}{\EX_P[F(T)] + F_P},
\end{equation}
where $F_P = \int P(T) \left(\frac{\partial \ln{P(T)}}{\partial T}\right)^2 dT$ is the Fisher information contained in the prior, and 
\begin{equation}
\EX_P[F(T)]=
\int P(T)dT\int P(\boldsymbol{X}|T) \left(\frac{\partial \ln{P(\boldsymbol{X}|T)}}{\partial T}\right)^2 d\boldsymbol{X},
\end{equation}
is the Fisher information of $P(\bm{X}|T)$, averaged over the prior. 
The VTSB, however, is generally not tight, unlike the CRB~\eqref{cramer_rao}. The reason is linked with the fact that since the  MSE \eqref{eq:FrequentistRisk1} scales with $1/nF(T)$ for $n$ large, as a consequence of Eq.\eqref{eq:FrequentistRisk2} the BMSE scales in the asymptotic limit as:
\begin{equation}\label{eq:asymptotic_mse}
\epsilon_{B}(\hat{T}(\boldsymbol{X}))
\sim
\EX_P\left[\frac{1}{n F(T)}\right]
\quad (n~\text{large}).
\end{equation}
In other words, the BMSE \eqref{eq:BayesianRisk1} scales with respect to $1/nF(T)$ averaged over the prior. Note that the VTSB \eqref{eq:VanTrees} on the other hand, scales with $1/\mathbb{E}_P[n F(T)]$ for $n$ large. Thus, by Jensen's Inequality \cite{Bickel2015, VanTrees2001} $\EX_P\left[1/F(T)\right] \geq 1/\EX_P[F(T)]$ and hence the bound is generally not tight.

It is also worth mentioning that the VTSB is not the only counterpart to the CRB. Other bounds may also take up this role \cite{Personick1971, Rubio2019}. Moreover, alternative bounds for Bayesian inference can be found in the literature \cite{Li2018, Liu2016, Lu2016, Weinstein2009, Tsang2012}. Results tailored for thermometry were recently obtained in~\cite{Rubio2020}, where bounds were derived to deal with said issues.
In addition, the authors also studied  concrete estimators and figures of merit, which include arguments on scale invariance, first put forth within thermometry in Ref.~\cite{Prosper1993}.

\subsection{\label{sec:numm}Efficient numerical Bayesian estimation}

Here we discuss a straightforward method of implementing Bayesian estimation numerically. 
More efficient methods may exist, but we have found  this approach to be both easy and efficient. It is also quite general, and can be readily extended to correlated outcomes (Appendix~\ref{app:corr}).
The goal is to compute the posterior~\eqref{eq:BayesTheorem} given a set of random $n$ outcomes $X_1,\ldots,X_n$.
Usually, one is also interested in assessing the results for increasingly larger sequences~\cite{Kiilerich2016,Gammelmark2013}.
There are  two main difficulties involved. 
First, dealing with the fact that $P(\bm{X}|T) = P(X_1|T)\ldots P(X_n|T)$ can be the product of a very large number of terms (and hence be very small); and second, the actual numerical computation of the normalization $P(\bm{X}) = \int  P(\bm{X}|T) P(T) dT$.

We handle both as follows. 
First, we discretize the temperature interval of interest, $[T_{\rm min}, T_{\rm max}]$, into $N_T$ points $T_k$,
so that the prior now becomes a discrete distribution $P_k$.
Second, we define the log-likelihood function 
\begin{equation}\label{loglike_discrete}
    L_{kn} = \sum\limits_{i=1}^n \ln P(X_i|T_k).
\end{equation}
This can be viewed as a matrix of size $N_T \times n$, which takes into account information up to time $n$. 
For instance, $L_{k,3}$ is using the information obtained from the $X_1,X_2,X_3$. 
Conveniently, $L_{kn} = L_{k,n-1} + \log P(X_n|T_k)$, so $L_{kn}$ can be constructed sequentially, by accumulating data from each new outcome. 
Eq.~\eqref{eq:BayesTheorem} may now be written as 
\begin{equation}\label{num_step1}
    P_{k|n} = \frac{e^{L_{kn}} P_k}{\sum_q e^{L_{qn}}P_q},
\end{equation}
where $P_{k|n}$ is a shorthand for $P(T_k|X_1\ldots X_n)$.

To stabilize the exponential, it is convenient to define the max of the log-likelihood, at each $n$, $L_{n}^{\rm max} = \max_k L_{kn}$.
We then rewrite Eq.~\eqref{num_step1} as 
\begin{equation}\label{num_step2}
    P_{k|n} = \frac{e^{L_{kn}-L_n^{\rm max}} P_k}{\sum_q e^{L_{qn}-L_n^{\rm max}}P_q}.
\end{equation}
This ensures that the most likely events will have the best numerical precision. 
We now see that Eq.~\eqref{num_step2} has the form 
\begin{equation}
    P_{k|n} = \frac{P_{kn}}{\sum\limits_q P_{qn}},
\end{equation}
where $P_{kn} = e^{L_{kn}-L_n^{\rm max}} P_k$ can be interpreted as a matrix of size $N_T \times n$, which is readily constructed from the matrix $L_{kn}$ and the vector $P_k$. 

It is now straightforward to compute any observable of interest. 
The BA~\eqref{eq:estimator}, for instance, becomes 
\begin{equation}\label{num_BA}
    \hat{T}_n 
    = \sum\limits_k T_k P_{k|n}.
    % = \frac{\sum\limits_k T_k P_{kn}}{\sum\limits_q P_{qn}}.
\end{equation}
The MSE~\eqref{eq:FrequentistRisk1} of a single realization $X_1,\ldots,X_n$ will then be $(\hat{T}_n - T_0)^2$, where $T_0$ is the true parameter.  Eq.~\eqref{eq:FrequentistRisk1} can be obtained by sample averaging this quantity over multiple realizations. 
The BMSE~\eqref{eq:BayesianRisk1}, on the other hand, is obtained in a similar way, but with data generated by randomly sampling temperatures from the prior. This step can be seen as a Monte Carlo integration.

% Conversely, the Bayesian error~\eqref{eq:BayesianRisk1}, associated to a single realization, is 
% \begin{equation}\label{num_MSE}
%     \epsilon(\hat{T}_n) = \sum\limits_{k}(T_k - \hat{T}_n)^2 P_{k|n}.
% \end{equation}
% This serves as a useful measure in itself, which is known as the posterior expected loss~\cite{Lehmann1998}. 
% However, sample averaging it over multiple realizations will not recover Eq.~\eqref{eq:BayesianRisk1}. To do that, one  must average data from samples obtained from different temperatures, taken from the prior distribution.

\section{\label{sec:analysis}Results}

\begin{figure*}
     \center
     \includegraphics[width=\textwidth]{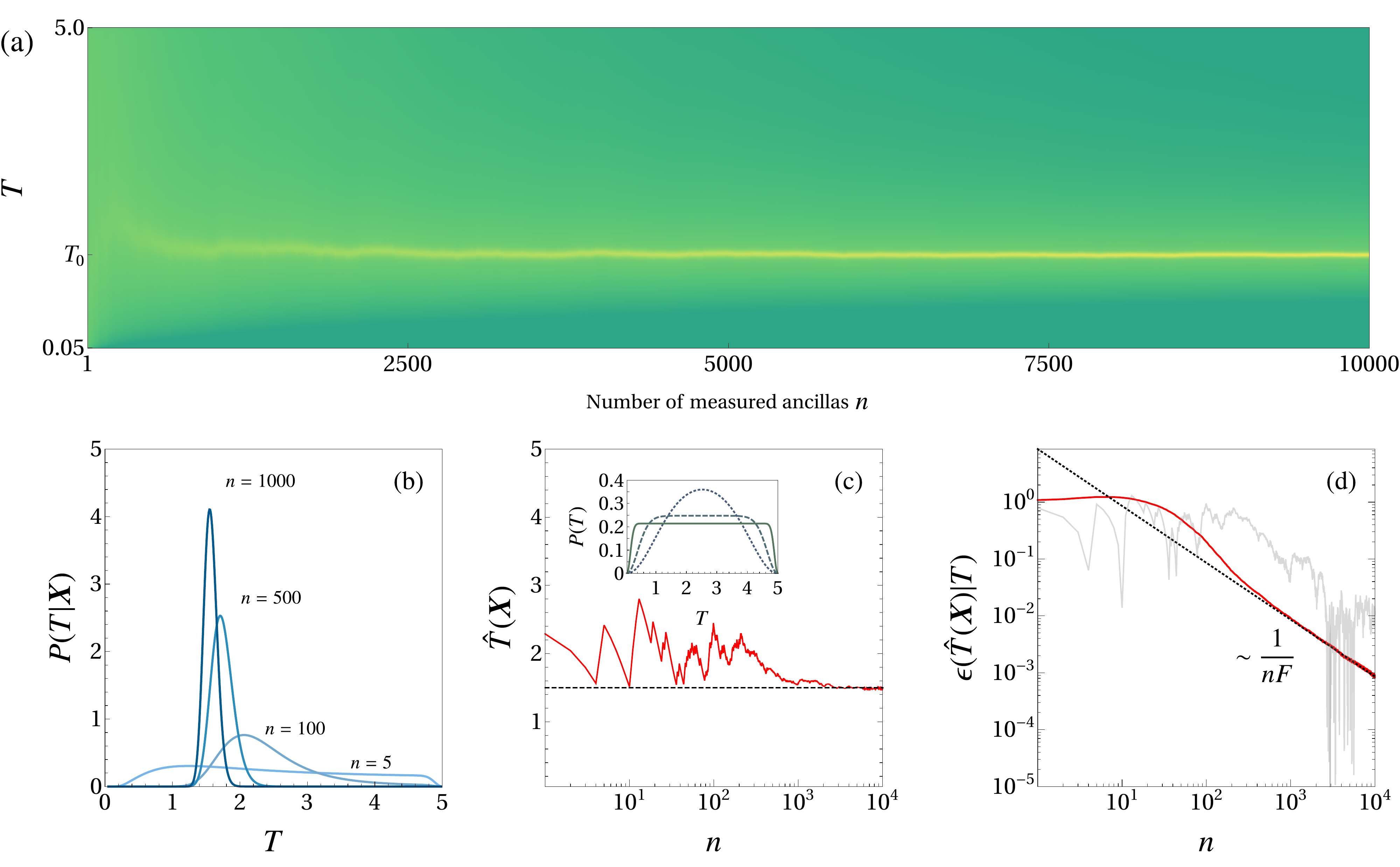}
     \caption{
     Bayesian estimation for collisional thermometry.
     The true temperature was chosen as  $T_0/\Omega = 1.5$.
     (a) Density plot of the posterior $P(T|X_1\ldots X_n)$ [Eq.~\eqref{num_step2}]. 
     (b) Same, but as a function of $T$, for select values of $n$. In both figures, the distribution is clearly seen to converge towards $T_0$ as $n$ increases.
     (c) Random realization of the BA~\eqref{eq:estimator}, as a function of $n$. 
     Inset: prior distribution~\eqref{eq:prior}, for  $\alpha = -1$ (dotted),  $\alpha = -10$ (dashed) and $\alpha = -100$ (solid).
     (d) MSE [Eq.~\eqref{eq:FrequentistRisk1}] for a single stochastic realization (gray), and averaged over multiple realizations (red). 
     For large $n$ it converges to $1/nF(T_0)$ (dotted), which saturates the CRB~\eqref{cramer_rao}.
     All curves were plotted using $\gamma \tau_{SE} = 0.4$, $g\tau_{SE} = \pi/2$ and $\alpha = -100$. The temperature was discretized in steps of $N_T = 500$, from $T_{\rm min} = 0.05$ to $T_{\rm max} = 5$.
     }
     \label{fig:PosteriorDensity}
\end{figure*}

We now reach the core results of this paper, where we implement the Bayesian estimation techniques of Sec.~\ref{sec:bayesian_estimation} to the collisional thermometry setting discussed in Sec.~\ref{sec:collisional_thermometry}.

As discussed in~\cite{Prosper1993,Rubio2020}, the choice of prior in thermometry is subtle, as it relates to the scale invariance of energy measurements. 
Since we are interested in generic measurements, generic energy spacings, and a non-equilibrium setting, we will take the prior for simplicity to be a
% The simplest choice of prior would be a
uniform distribution over a certain range $[T_{\rm min}, T_{\rm max}]$. 
% Here we take instead,the 
Or, what is
slightly more general 
% distribution~
\cite{Li2018}
\begin{equation}\label{eq:prior}
P(T)=\frac{1}{(T_{max}-T_{min})}
\lambda_{\alpha} \left(\frac{T-T_{min}}{T_{max}-T_{min}}\right)
\end{equation}
where
\begin{equation}
\lambda_{\alpha}(\theta) = \frac{e^{\alpha  \sin ^2(\pi  \theta  )}-1}{e^{\alpha /2} I_0\left(\frac{\alpha }{2}\right)-1},
\end{equation} 
and $I_0$ is the modified Bessel function of the first kind. 
This is plotted in the inset in Fig.\ref{fig:PosteriorDensity} (c), for different values of $\alpha$.
It is sharply peaked for $\alpha > 0$, and tends to a smoothed  uniform  when $\alpha$ is negative and large. 
It thus allows us to conveniently interpolate between a sharply peaked distribution, and a flat one, while preserving the (possibly physical) constraint that the temperature should lie within a specific interval. From hereafter we will perform all simulations considering $\alpha = -100$. Another advantage of this prior concerns the VTSB~\eqref{eq:VanTrees}, which does not hold for truncated distributions, like the uniform \cite{bj/1186078362, Ramakrishna2020}.

% \goa{
% We also point out, however, that this choice of prior and estimators, while simple, is not optimal in the presence of limited data. This regime is properly handled when one performs scale-invariant estimation, since, as we will see,  all the quantities here depend only on the ratio $\Omega/T$. This consideration results in a logarithmic error and a prior of the form $P(T) \propto 1/T$, as it was shown in \cite{Rubio2020} (for a discussion in classical thermometry, see \cite{Prosper1993}).
% }

Basic results are summarized in Fig.~\ref{fig:PosteriorDensity}. 
For a fixed $T_0$, we generate a sequence of random outcomes $X_i$ from $P(X_i|T)$ in Eq.~\eqref{eq:CollisionalLikelihood}. 
In Fig.~\ref{fig:PosteriorDensity}(a) we show the posterior distribution $P(T|X_1\ldots X_n)$ [Eq.~\eqref{num_step2}], with the vertical axis representing the temperature, and the horizontal axis the number of measured ancillas. This presentation of the Bayesian updating scheme and the posterior distribution was strongly motivated by the seminal works in \cite{Kiilerich2016} and \cite{Gammelmark2013}.
Fig.~\ref{fig:PosteriorDensity}(b) plots the same results, but as a function of $T$.
As can be seen in both images, the posterior is initially broad for  few outcomes, but gradually improves with increasing $n$. 
For $n=1000$ it is already sharply peaked around $T_0$. 
And, as predicted by~\eqref{posterior_gaussian}, the precision continues to improve with increasing $n$, with the variance scaling as $1/nF(T_0)$.

The BA is shown in Fig.~\ref{fig:PosteriorDensity}(c).
It is noisy up to $n\sim 1000$, but then quickly converges towards the true value $T_0$. 
The error associated to the realization of $\hat{T}(\bm{X})$ in Fig.~\ref{fig:PosteriorDensity}(c), is plotted as a gray curve in Fig.~\ref{fig:PosteriorDensity}(d). 
Overall, it oscillates significantly, but gradually tends to zero (notice the log scale). 
Averaging this over multiple realizations $\bm{X}$, yields the MSE~\eqref{eq:FrequentistRisk1}, which is plotted by the solid red line in Fig.~\ref{fig:PosteriorDensity}(d).
As can be seen, in the large $n$ limit, it converges to $1/nF(T_0)$, shown by the dotted line. It hence saturates the CRB~\eqref{cramer_rao}.

The results of Fig.~\ref{fig:PosteriorDensity} refer to the full swap between system and ancilla, $g\tau_{SA} = \pi/2$. In this case, the distribution $P(X_i|T)$ in Eq.~\eqref{eq:CollisionalLikelihood}, can be easily computed analytically and acquires the particularly simple form
\begin{equation}\label{eq:collisional_likelihood}
P(X_i=1|T) = 
\frac{1-e^{-\Gamma}}{1 + e^{\frac{\Omega}{T}}}
\end{equation}
where $\Gamma = \gamma (2 \bar{n} +1) \tau_{SE}$ is the thermal relaxation parameter. As it was shown in \cite{Seah2019}, for the regime in consideration the FI also acquires a tractable form. One may simply use the likelihood given in Eq.~\eqref{eq:collisional_likelihood} above, together with \eqref{eq:fisher} to find: %
\begin{equation}\label{eq:collisional_fisher}
\frac{F}{\mathcal{F}_{th}} 
=
\frac{(\bar{n}+1) \left(e^{\Gamma}+2\bar{n} \Gamma -1\right)^2}{e^{2 \Gamma } (\bar{n}+1)-e^{\Gamma }-\bar{n}}.
\end{equation}
And as it was pointed out, the dependence on $\Gamma$, which would not be present in a fully thermalized ancilla,  is responsible for the enhancement over the thermal precision. 
As a consequence, the error in Fig.~\ref{fig:PosteriorDensity}(d) actually surpasses the precision of the thermal Fisher information~\eqref{qubit_TFI}.

\begin{figure}
     \center
     \includegraphics[width=\columnwidth]{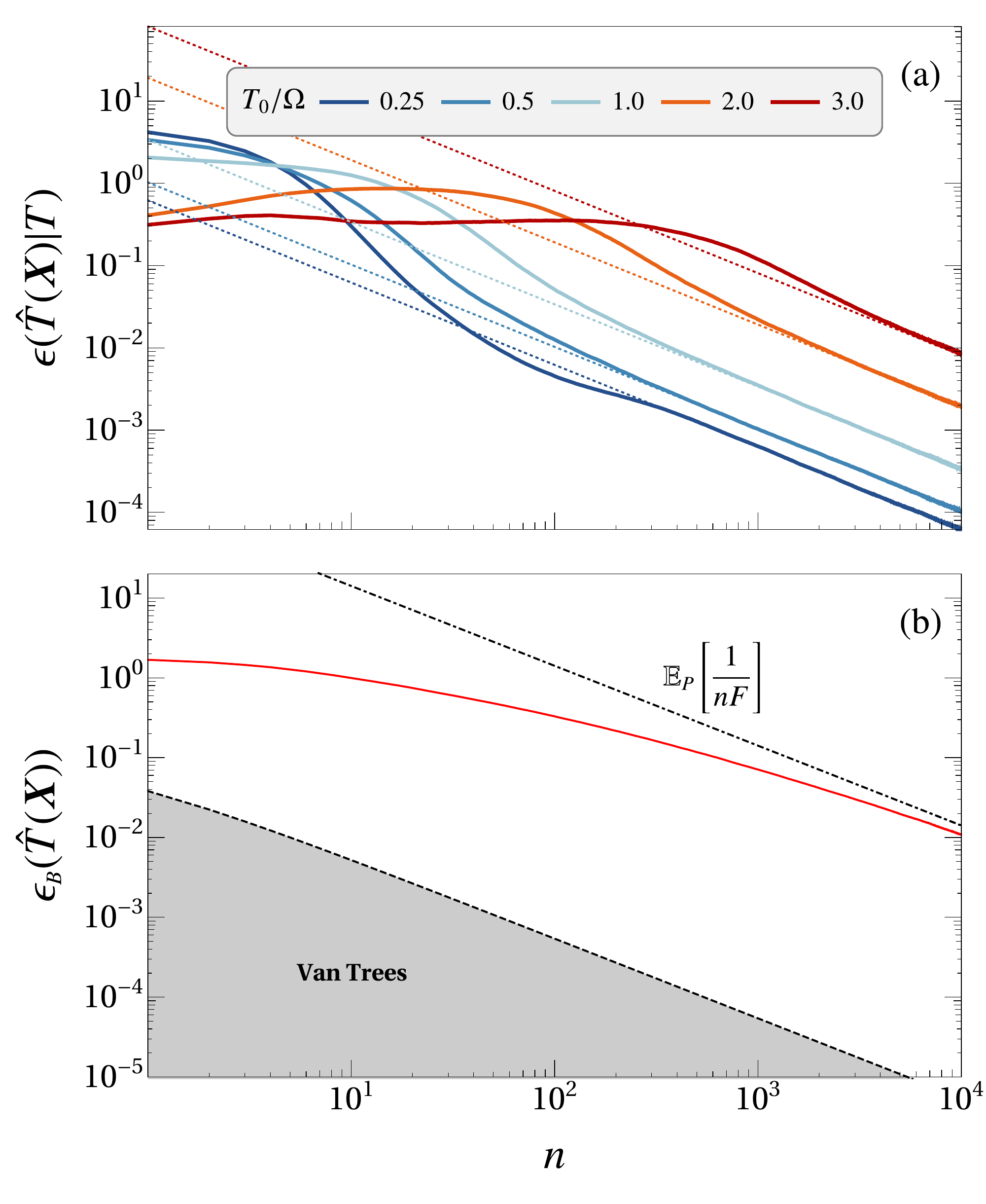}
     \caption{(a) We plot the MSE \eqref{eq:FrequentistRisk1} for different values of $T$ averaged numerically over 3000 different trajectories. The dotted lines correspond to the CRB for the different temperatures shown in the legend. (b)  We repeat the procedure in (a), averaging the MSE over 500 trajectories and also over the prior distribution \eqref{eq:prior}, obtaining the BMSE [Eq.~\eqref{eq:BayesianRisk1}]. The integral over the prior is also performed numerically through a temperature discretization with $N_T = 150$. We also show the asymptotic limit defined in \eqref{eq:asymptotic_mse} (dot-dashed).  Other parameters are the same as in Fig.~\ref{fig:PosteriorDensity}.
     }
     \label{fig:MSEPlot}
\end{figure}

The MSE, similar to Fig.~\ref{fig:PosteriorDensity}(d),  is plotted in Fig.~\ref{fig:MSEPlot}(a) for different values of $T_0$. 
The dashed lines in all cases refer to the asymptotic limit $1/nF(T_0)$. 
We can see that the estimation is more accurate as the temperature decreases, which is attributed to the larger sensitivity on $T$ in the likelihood Eq.~\eqref{eq:CollisionalLikelihood}.
As argued in Sec.~\ref{sec:bayesian_estimation}, the MSE in Fig.~\ref{fig:MSEPlot}(a) depends on the actual value of $T_0$, which is not known. 
Hence, it is convenient to analyze the BMSE $\epsilon_B$ from Eq.~\eqref{eq:BayesianRisk1}.
This is shown in Fig.~\ref{fig:MSEPlot}(b). 
It quantifies the overall expected performance of the estimator, for the temperature range $[T_{\rm min},T_{\rm max}]$.
This curve is bounded by the VTSB~\eqref{eq:VanTrees}, denoted by the gray region in Fig.~\ref{fig:MSEPlot}(b). Hence, as can be seen, in this example the bound is still quite loose, for reasons pointed out at the end of Sec.~\ref{subsec:BayesianEstimators}. 
As portrayed in Fig.~\ref{fig:MSEPlot}(b), note how the BMSE converges to the asymptotic limit in Eq.~\eqref{eq:asymptotic_mse} instead. We can see that in order to investigate its asymptotic behavior, it suffices to calculate the usual CRB averaged over the Prior \eqref{eq:asymptotic_mse}, with the Fisher Information given by \eqref{eq:fisher}.

\begin{figure*}
     \center
     \includegraphics[width=\textwidth]{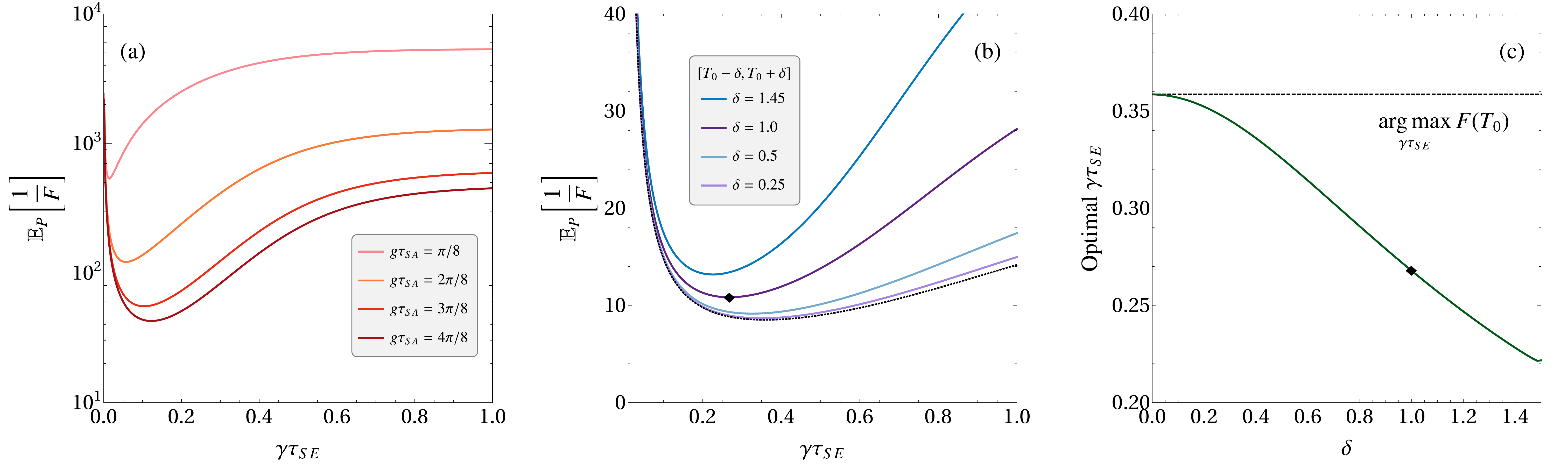}
     \caption{
    (a) We plot the expectation in the RHS of Eq.~\eqref{eq:asymptotic_mse} as a function of $\gamma \tau_{SE}$ considering the same temperature range $[T_{min}, T_{max}]$ from Fig.~\ref{fig:PosteriorDensity}.
    (b) We fix the SA coupling in the full swap regime $g\tau_{SA} = \pi/2$ and plot Eq.~\eqref{eq:asymptotic_mse} for different temperature intervals. Note how the minima shifts to the left as $\delta$ increases. (c) We plot the optimal value of $\gamma \tau_{SE}$ as we increase the size of the interval $[T_0 - \delta, T_0 + \delta]$. We also show the optimal value (dashed) for $F(T_0)$ at the particular temperature $T_0/\Omega = 1.5$. The diamond symbol hightlights the minimum for $\delta = 1$. The prior used here is also given by Eq.~\eqref{eq:prior}, with $\alpha=-100$, but the endpoints are changed as described above. 
     }
     \label{fig:optimization_intervals}
\end{figure*}

Thus, in Fig.~\ref{fig:optimization_intervals} we turn to the BMSE \eqref{eq:BayesianRisk1} and its asymptotic value \eqref{eq:asymptotic_mse} in order to investigate this effect more systematically. 
In Fig.~\ref{fig:optimization_intervals}(a) we plot Eq.~\eqref{eq:asymptotic_mse} for different values of $g \tau_{SA}$ as a function of $\gamma \tau_{SE}$. Note that this plot is actually independent of the true temperature, but depends only on the choice of temperature interval. 
Hence,  it provides a general view on how the choice of parameters affects the  asymptotic performance of the protocol. The smaller the value of $\mathbb{E}_P[1/nF]$ the better the estimation. 
Therefore, in Fig.~\ref{fig:optimization_intervals}(a) we can see how the asymptotic error and the optimal value of $\gamma \tau_{SE}$ depends on the effective SA coupling $g \tau_{SA}$.

On a similar note, we investigate how the optimal choice of parameters may change depending on the temperature interval of the prior. 
In Fig.~\ref{fig:optimization_intervals}(b) we plot $\EX_P[1/F]$ as a function of $\gamma \tau_{SE}$. 
This time around we consider a symmetric interval
from $T_{\rm min} = T_0 - \delta$ to $T_{\rm max} = T_0 + \delta$, centered at $T_0/\Omega = 1.5$ for different values of $\delta$. We can see that the optimal choice of $\gamma\tau_{SE}$ clearly depends on the temperature interval in consideration. In particular, we can verify from this plot that for larger intervals, the optimal regime is narrower, and the error quickly increases with $\gamma \tau_{SE}$. Conversely, an increase on the temperature interval requires a decrease in $\gamma \tau_{SE}$ in order to achieve optimality.  Moreover, we can also see that as the interval narrows, both the asymptotic error and the optimal SE coupling coincide with the results found for the temperature $T_0$. Finally, note from Fig.~\ref{fig:optimization_intervals}(c) how the optimal parameters continuously decrease as one increases the temperature range.

This analysis shows how the BMSE is particularly useful in the search for optimal parameters to enhance precision. 
The Fisher information and the CRB~\eqref{cramer_rao} depend on the actual temperature. 
Thus,  the values of $\gamma\tau_{SE}$ and $g\tau_{SA}$ which are optimal for a given $T$, are not necessarily optimal for another. 
And since the true value of $T$ is not known, this introduces a conundrum. 
Bayesian estimation avoids this by focusing on an entire range of temperatures, quantified by the prior $P(T)$. 
By focusing on the asymptotic BMSE ($n\to \infty$), as compared to the asymptotic MSE $1/nF(T)$, in Fig.~\ref{fig:optimization_intervals} we showed how the BMSE in Fig.~\ref{fig:MSEPlot}(b) can be optimized over $\gamma\tau_{SE}$ and $g\tau_{SA}$, to yield a strategy which is good for the entire temperature range.

\section{\label{sec:discussions_and_conclusions}Discussions and conclusions}

In this paper we have put forth a concrete estimation protocol based on the collisional thermometry setup proposed in \cite{Seah2019}, showcasing how the Bayesian framework may display further insights as a thermometric tool, providing a simple alternative to easily process the data. Bayes theorem was used to sequentially update the temperature distribution, updated on the measurement outcomes. The performance of the estimators were then assessed through the Bayesian MSE. These results were then compared with the van Trees-Sch\"utzenberger inequality, a Bayesian counterpart of the Cramér-Rao bound. Finally, by investigating the Bayesian MSE in the asymptotic limit we also showed how it can be used to perform an analysis of the model which is independent of the temperature. By doing so we were able to find the optimal parameters for the model, minimizing the BMSE in the asymptotic limit.

In principle, it's also possible to further generalize the protocol here for collective measurement on the ancillas, investigating how correlations affect the estimations. This also further enriches the discussion on how to choose the measurement basis, since it may acquired a more sophisticated form, assuming a dependence on the temperature. A possible alternative would be, for instance, to employ adaptive strategies \cite{Escher2011}.

In a more general picture, we have only scratched the surface of what Bayesian estimation offers. Further research directions could go into direction of investigating other estimators, aiming into uncovering different estimation protocols and estimators under other regimes or prior distributions. While of limited purpose here, minimax estimators are such an example \cite{Lehmann1998}. Even the choice of a prior distribution may not be entirely straightforward and must be carefully investigated \cite{Jaynes2003, VonToussaint2011}.

Finally, further work can be done on the generalization of a few well-known concepts in both quantum thermometry and also quantum metrology in general, such as it was done in \cite{Martinez-Vargas2017, Demkowicz-Dobrzanski2020}. We clarify here, however, that global treatments are in no way exclusive to the frequentist approach, as it's always possible to construct a global Bayesian framework for the estimation problem (see e.g \cite{Personick1971, helstrom1976, Holevo2011} for a fully Bayesian treatment). In the same manner, the frequentist approach is just as useful when considering concrete protocols. Both approaches are not mutually exclusive, but rather, the focus on how they are used just shifts depending on the problem at hand. We also stress that the tools presented here are in no way restricted to thermometry. As it was shown by many of the works cited here, Bayesian estimation has been successfully employed in the quantum metrology community in several different contexts, albeit relatively few and far between in thermometry.

\vspace{0.1cm}

\section*{Acknowledgments}

GOA acknowledges the financial support from the S\~ao Paulo founding agency FAPESP (Grant No. 2020/16050-0) and CAPES.
GTL acknowledges the financial support of the S\~ao Paulo Funding Agency FAPESP (Grants No.~2017/50304-7, 2017/07973-5 and 2018/12813-0), the Eichenwald foundation (Grant No.~0118 999 881 999 119 7253), and the Brazilian funding agency CNPq (Grant No. INCT-IQ 246569/2014-0). 

\appendix
\section{\label{app:corr}Bayesian inference for correlated ancillas}

As a proof of principle, we have focused on the case where the collisional thermometry outcomes can be taken to be approximately independent [Eq.~\eqref{eq:DistributionApproximation}]. 
But the framework is not restricted to this case. 
More generally, starting from a joint distribution~\eqref{likelihood_geral}, we can decompose 
\begin{widetext}
\begin{equation}
    P(X_1,\ldots, X_n|T) = P(X_n|T,X_1,\ldots,X_{n-1})~P(X_{n-1} | T, X_1, \ldots, X_{n-2})\ldots~P(X_2|T,X_1)~P(X_1|T).
\end{equation}
\end{widetext}
Since collisional thermometry yields a well defined causal order in the outcomes, these transition probabilities can all be directly obtained from the model. 
Focusing on the case where $T$ is discretized in steps $T_k$, we can now generalize Eq.~\eqref{loglike_discrete} to 
\begin{equation}\label{loglike_hierarchy}
    L_{kn} = \sum\limits_{i=1}^n \ln p(X_i|T_k,X_1,\ldots,X_{i-1}).
\end{equation}
With this small modification, all other results in the paper continue to be valid, even in the case of dependent outcomes. 
That is to say, the formalism itself does not change; all that changes is how we construct the likelihood. 
This is quite remarkable, and very nice attribute of BE.

The results of Fig.~\ref{fig:mutualinfodecay} show that the mutual information always decays with the distance between the ancillas. 
Hence, in practice, one does not need to retain the full hierarchy of distributions in Eq.~\eqref{loglike_hierarchy}. 
Instead, one may truncate it at a given Markov order. 
For instance, assuming that only nearest-neighbor correlations are important, one may approximate 
\begin{equation}\label{loglike_mark}
    L_{kn} \simeq \sum\limits_{i=1}^n \ln P(X_i|T_k,X_{i-1}),
\end{equation}
where $P(X_i|T_k,X_{i-1})$ forms essentially a Markov chain. 
Or one may consider two neighbors, $P(X_i|T_k,X_{i-1},X_{i-2})$, and so on. 
This can be very useful because, in practice, constructing the theoretical model $p(X_i|T_k,X_1,\ldots,X_{i-1})$ for a large number of ancillas is hard due to the increasing dimension of the global Hilbert space.
A distribution such as $p(X_i|T_k,X_{i-1})$, on the other hand, depends only on two ancillas, and hence is analytically/numerically  manageable. 

\begin{figure}
     \center
     \includegraphics[width=\columnwidth]{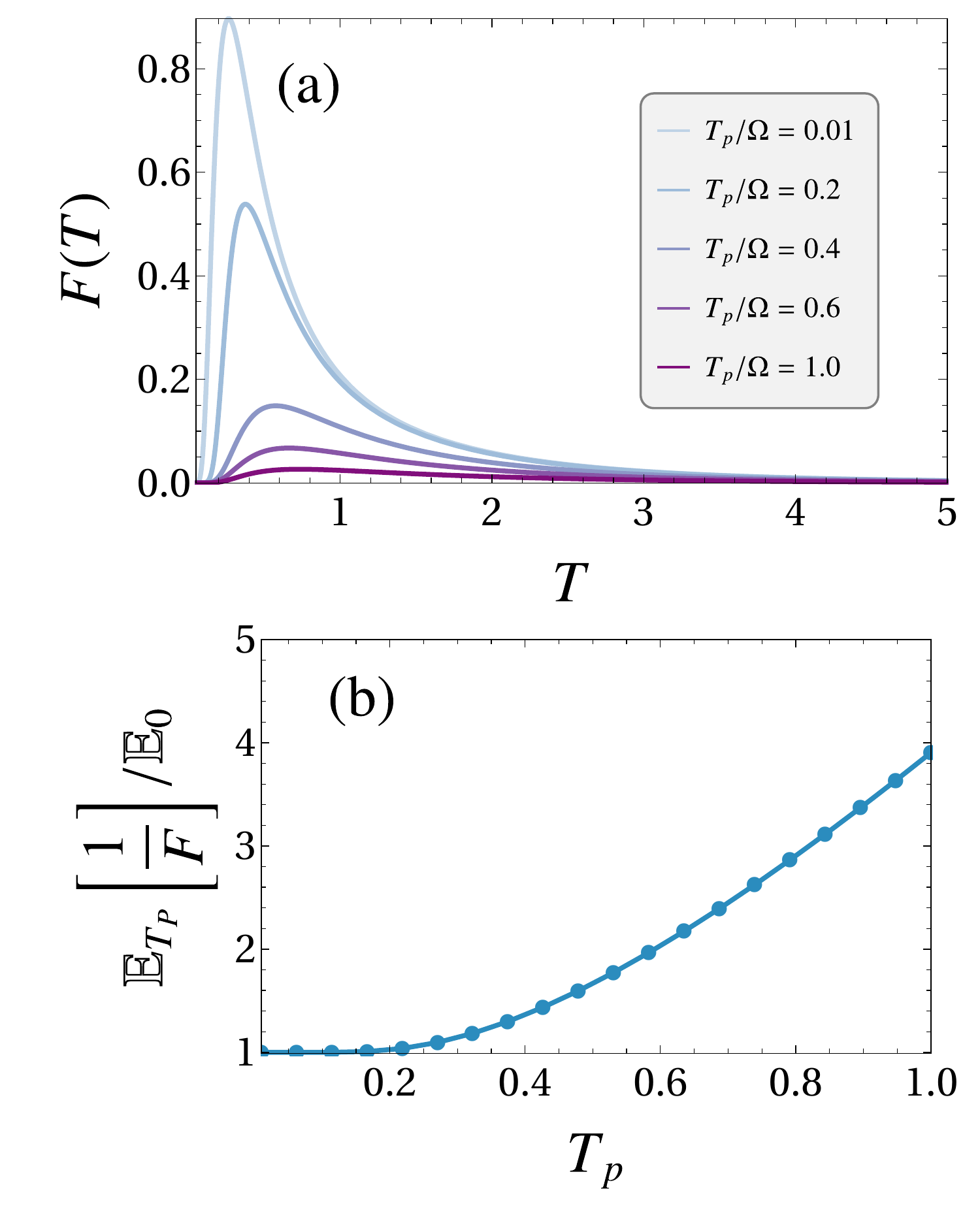}
     \caption{(a) We plot the Fisher information of the likelihood in Eq.~\eqref{eq:collisional_likelihood_probe} for different values of $T_p$. (b) We calculate the ratio between the asymptotic Bayesian risk~\eqref{eq:asymptotic_mse} obtained by integrating the FI in (a), from $T_{min} = 0.1$ to $T_{max} = 5$ and the asymptotic Bayesian risk for $T_p = 0$. All the other parameters are the same as in Fig. ~\ref{fig:PosteriorDensity}.
     }
     \label{fig:ProbeErrorPlot}
\end{figure}

\section{\label{app:probe}Effect of noisy probes}

In the main text we assumed an ideal scenario where one can always initialize the ancillas in the desired state, namely the ground-state. Here we further generalize our approach for a situation where the observer does not have perfect control over the probe states.

Assuming that the ancillas are initialized in a thermal state, we first investigate how the temperature of the probes affect the asymptotic precision of the protocol, which is related to Eq.~\eqref{eq:asymptotic_mse}. Instead of Eq.~\eqref{eq:collisional_likelihood}, the likelihood assumes the form
\begin{equation}\label{eq:collisional_likelihood_probe}
P_{T_P}(X_i=1|T) = 
\frac{e^{-\Gamma}}{1+e^{\frac{\Omega}{T_p}}}+
\frac{1-e^{-\Gamma}}{1 + e^{\frac{\Omega}{T}}},
\end{equation}
instead. This result is a consequence of the linearity of the stroboscopic map from Eq.~\eqref{eq:stroboscopic_map}; the resulting likelihood for the thermalized probe is simply a convex combination of the resulting likelihood for ancillas initialized in the states $|0\rangle\langle 0|$ and $|1\rangle\langle 1|$, weighted by the Gibbs probabilities. Additionally, as a consequence of the convexity of the FI \cite{Cohen1968}, the resulting precision will be smaller than what one would get for an ancilla initialized in the ground state.

In particular, we are interested in the asymptotic value of the Bayesian error given by Eq.~\eqref{eq:asymptotic_mse}. To perform comparisons with the ideal case, we first write the asymptotic error for ground-state ancillas as $\mathbb{E}_0$, which can be calculated from the FI in Eq.~\eqref{eq:collisional_fisher}. In Fig.~\ref{fig:ProbeErrorPlot}(a) we show the Fisher information $F(T)$ for different values of $T_p$. In Fig.~\ref{fig:ProbeErrorPlot}(b) we show how much precision is lost when compared to the case where $T_p = 0$, i.e. we plot the ratio between $\mathbb{E}_{T_p}[1/F]$ and $\mathbb{E}_0$ for different probe temperatures.
\begin{figure}
     \center
     \includegraphics[width=\columnwidth]{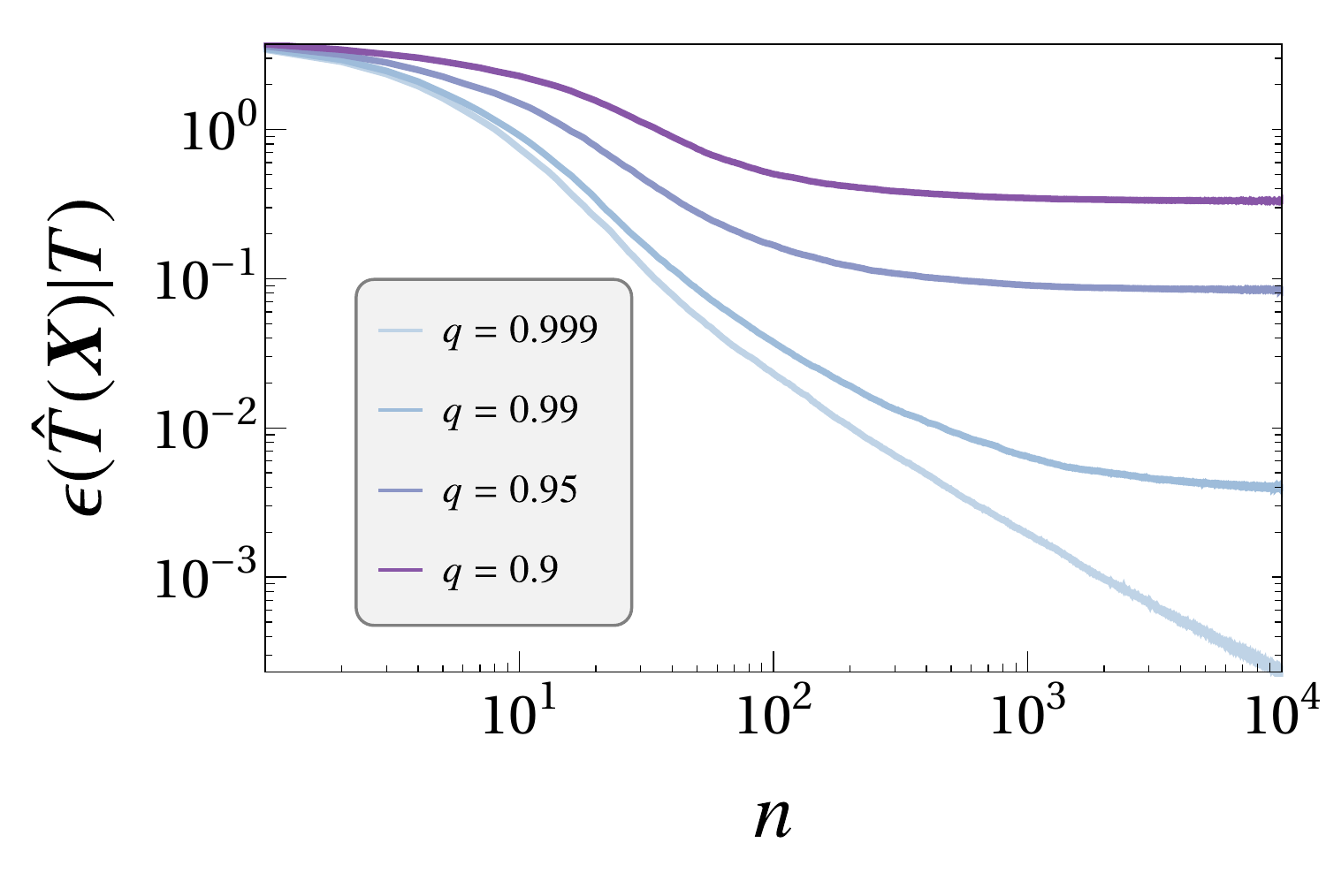}
     \caption{Bayesian MSE~\eqref{eq:BayesianRisk1} calculated for outcomes generated from the likelihood in Eq.~\eqref{eq:collisional_likelihood_bias} for different values of $q$.  This type of bias introduces a systematic error; the Bayesian risk initially decreases with a $1/n$ scaling but eventually saturates, since the estimation converges to a wrong value of temperature. All the other parameters in the simulation are the same as in Fig.~\ref{fig:ProbeErrorPlot}.
     }
     \label{fig:ProbeErrorVanTreesPlot}
\end{figure}

Now, we also investigate a second scenario: we are interested in what happens when ancillas are prepared in the states $|0\rangle\langle 0|$ and $|1\rangle\langle 1|$ with probabilities $q$ and $1-q$, respectively, but the observer has no access to these probabilities. In other words, the outcomes are generated from the likelihood
\begin{equation}\label{eq:collisional_likelihood_bias}
P(X_i|T) = 
q P(X_i|T, \rho_{A, 0})
+
(1-q)  P(X_i|T, \rho_{A, 1}),
\end{equation}
where $\rho_{A, k} = | k \rangle \langle k |$, with $k=0,1$. The inference however is still performed with respect to the ideal model in~\eqref{eq:collisional_likelihood}. As we show in Fig.~\ref{fig:ProbeErrorVanTreesPlot}, this introduces a persistent error into the estimation. Since the experimenter is using an incorrect model for the likelihood, the resulting estimation will deviate from the true value of the temperature. The saturated error will correspond to the difference between the true value $T_0$ of the temperature and the temperature one would get from the ideal likelihood~\eqref{eq:collisional_likelihood} for the given detection record.

\bibliography{library}

\end{document}